%% file: main.tex
\documentclass{vgtc}                          % final (conference style)
\ifpdf%                                % if we use pdflatex
  \pdfoutput=1\relax                   % create PDFs from pdfLaTeX
  \usepackage{graphicx}                % allow us to embed graphics files
  \DeclareGraphicsExtensions{.pdf,.png,.jpg,.jpeg} % for pdflatex we expect .pdf, .png, or .jpg files
\else%                                 % else we use pure latex
  \usepackage{graphicx}                % allow us to embed graphics files
  \DeclareGraphicsExtensions{.eps}     % for pure latex we expect eps files
\fi%

\graphicspath{{figures/}{pictures/}{images/}{./}} % where to search for the images

\usepackage{microtype}                 % use micro-typography (slightly more compact, better to read)
\PassOptionsToPackage{warn}{textcomp}  % to address font issues with \textrightarrow
\usepackage{textcomp}                  % use better special symbols
\usepackage{mathptmx}                  % use matching math font
\usepackage{times}                     % we use Times as the main font
\usepackage{cite}                      % needed to automatically sort the references
\usepackage{tabu}                      % only used for the table example
\usepackage{booktabs}                  % only used for the table example
\usepackage{amsmath}
\usepackage{amssymb}
\usepackage{enumitem}
\usepackage{textcomp}
\usepackage{multirow}
\usepackage[table,xcdraw]{xcolor}
\onlineid{1299}

\vgtccategory{Research}

\vgtcinsertpkg

\title{\textit{FVA:}
Modeling Perceived Friendliness of Virtual Agents \\ Using Movement Characteristics}

\author{Tanmay Randhavane\thanks{e-mail: tanmay@cs.unc.edu}\\ %
        \parbox{2.0in}{\scriptsize \centering Dept of Computer Science \\ University of North Carolina at Chapel Hill}%
\and Aniket Bera\thanks{e-mail: ab@cs.unc.edu}\\ %
     \parbox{2.0in}{\scriptsize \centering Dept of Computer Science \\ University of North Carolina at Chapel Hill}%
\and Kyra Kapsaskis\thanks{e-mail: kyrakaps@gmail.com}\\ %
     \parbox{2.0in}{\scriptsize \centering Dept of Psychology and Neuroscience \\ University of North Carolina at Chapel Hill}%
\and Kurt Gray\thanks{e-mail: kurtjgray@gmail.com}\\ %
     \parbox{2.0in}{\scriptsize \centering Dept of Psychology and Neuroscience \\ University of North Carolina at Chapel Hill}%
\and Dinesh Manocha\thanks{e-mail: dm@cs.umd.edu}\\ %
     \parbox{2.0in}{\scriptsize \centering Dept of Computer Science and Electrical \& Computer Engineering \\ University of Maryland at College Park}}

\teaser{
   \includegraphics[width=\linewidth]{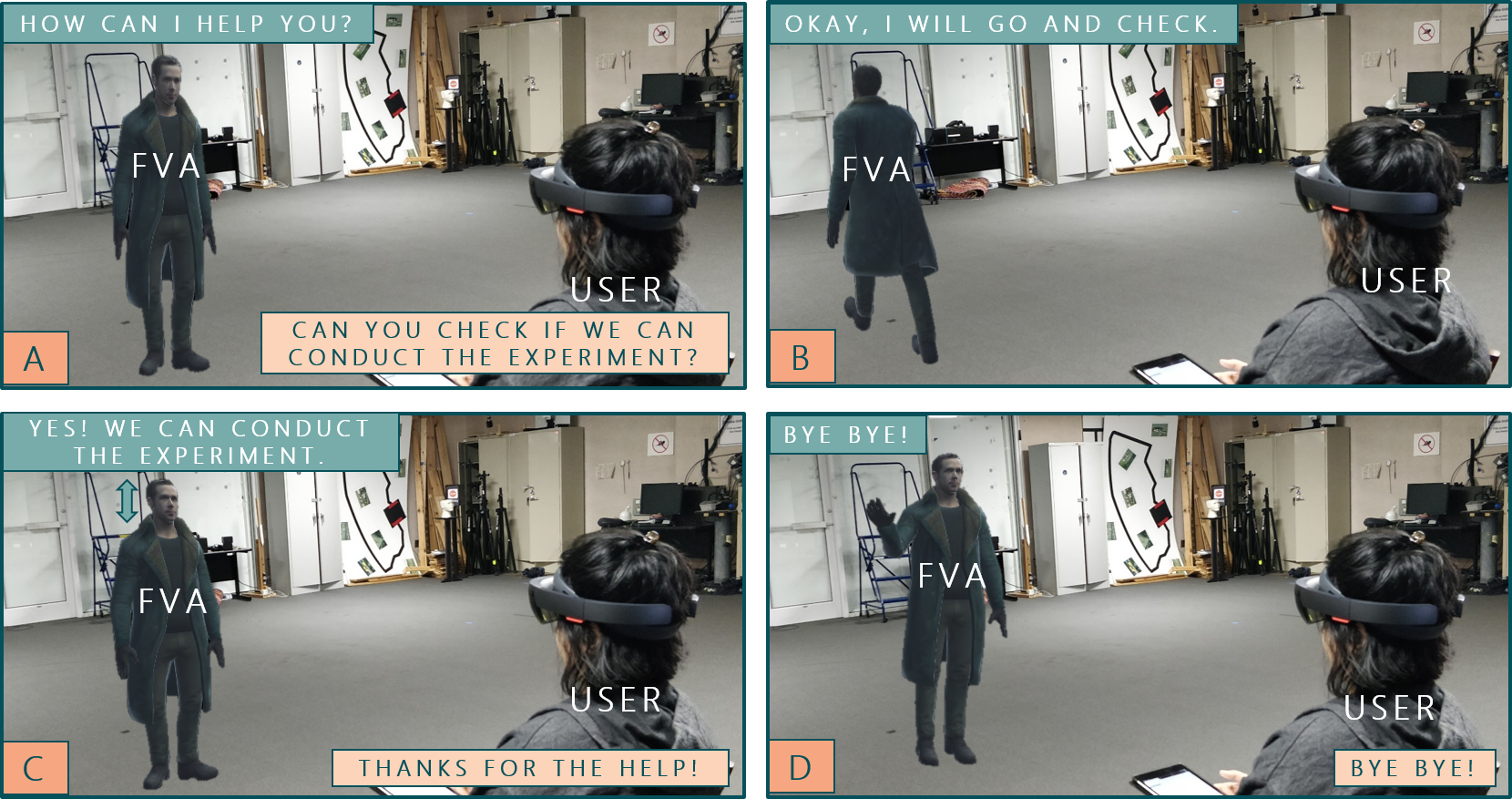}
   \caption{\textbf{Friendly Virtual Agent (FVA)}: We present an algorithm to model perceived friendliness of a virtual agent by varying its gaze (A), gait (B), and gestures corresponding to head nodding (C) and waving (D). These movement cues are generated using our novel data-driven friendliness model. We evaluate the benefits in terms of an improved sense of social presence using an AR validation study using a HoloLens.
   }
   \label{fig:cover}
}

\abstract{
We present a new approach for improving the friendliness and warmth of a virtual agent in an AR environment by generating appropriate movement characteristics. Our algorithm is based on a novel data-driven friendliness model that is computed using a user-study and psychological characteristics. We use our model to control the movements corresponding to the gaits, gestures, and gazing of friendly virtual agents (FVAs) as they interact with the user's avatar and other agents in the environment. We have integrated  FVA agents with an AR environment using with a Microsoft HoloLens. Our algorithm can generate plausible movements at interactive rates to increase the social presence.  We also investigate the perception of a user in an AR setting and observe that an FVA has a statistically significant improvement in terms of the perceived friendliness and social presence of a user compared to an agent without the friendliness modeling. We observe an increment of $5.71\%$ in the mean responses to a friendliness measure and an improvement of $4.03\%$ in the mean responses to a social presence measure.
} % end of abstract

\begin{document}



\maketitle

\input{1_introduction.tex}

\input{2_related.tex}
\input{3_overview.tex}

\input{4_FriendlinessModel.tex}
\input{5_FVAs.tex}
\input{6_userStudy.tex}

\input{7_Conclusion.tex}

\acknowledgments{This research is supported in part by ARO grant W911NF-19-1- 0069, Alibaba Innovative Research (AIR) program and Intel.}

\bibliographystyle{abbrv-doi}

\bibliography{template}
\end{document}

%% file: 1_introduction.tex
\section{Introduction} % 1.5 pages

Intelligent Virtual Agents (IVAs) corresponding to embodied digital or virtual characters are widely used in augmented or mixed reality environments. In many applications corresponding to treatment and rehabilitation, virtual assistance, and training, IVAs can help the user accomplish a variety of goals. Moreover, many of these applications demand that the generated IVAs look, move, behave, communicate, or act like ``living" creatures, whether real or fictional.

There is considerable work on 3D modeling and generating realistic appearances of IVAs using a combination of capturing, computer vision, and rendering techniques. The resulting algorithms are being integrated with current game engines and AR systems~\cite{li2019technical,facebook}. At the same time, a major issue with IVAs is that they are expected to be believable and exhibit behaviors and emotions that make them appear plausible in an AR environment. In particular, the user should feel social presence with the IVAs and should feel confident in their abilities to perform the designated tasks~\cite{bente2008avatar,kim2017exploring,kim2018does}. According to prior literature~\cite{youngblut2003experience}, ``social presence occurs when users feel that a form, behavior, or sensory experience indicates the presence of another individual. The amount of social presence is the degree to which a user feels access to the intelligence, intentions, and sensory impressions of another". It has been shown that the IVAs that exhibit social behavior and human-like personality traits can improve the sense of social presence~\cite{Romano2005BASICAB}.

%, i.e., to have a consistent behavior, to exhibit some form of personality and emotions, to communicate and interact in a plausible way, 
%A distinctive characteristic of IVAs is that their behavior should exhibit some aspects of human intelligence, including autonomous behavior, communication and coordination with other IVAs, dialogues with humans, and learning capabilities.
%For many AR applications such as treatment and rehabilitation, virtual assistance, gaming, etc., Intelligent Virtual Agents (IVAs) can be used to help the user to accomplish a variety of goals.
%These applications require that the 
%
%When these IVAs are introduced into the physical world, the user should feel that these IVAs have a sense of awareness and can influence the user and objects in the physical world.
 %should feel that these virtual agents have an awareness of and can influence the real world. For example, an IoT connected lamp that a digital assistant such as Amazon Alexa can turn on or off increases a user's confidence in Alexa's capabilities~\cite{kim2018does}. 
 %This confidence is also affected by the social behavior of IVAs and non-verbal cues~\cite{kim2015maintaining}.
\nocite{kim2015maintaining}

% . By presenting nonverbal cues, these virtual agents can communicate their intentions and goals to the user, thereby increasing the user's confidence in them~.

The challenge of how to make a user trust an IVA is mirrored by an analogous issue in psychology -- how to get people to trust and rely more on other people in joint tasks. When people interact with others in joint tasks, they evaluate them on two broad issues: competence and warmth/friendliness~\cite{fiske2007universal}. For AR applications, competence is related to an IVA's ability and intelligence and can be increased by giving an IVA the ability to sense and affect the real world (e.g., using voice, camera, or depth sensors). Although competence is an important consideration when evaluating potential action partners, an even more important consideration in terms of evaluating others' is warmth~\cite{Bergmann2012Second}. Warmth is one of the most fundamental interpersonal valuations and deals with whether a person is friendly or not~\cite{argyle1998bodily}. Even if a person is competent, if they do not exhibit friendliness, then people don't trust them and are not likely to choose them as an interaction partner. Similarly, users are likely to have more confidence in IVAs that appear friendlier. 
%Therefore when designing IVAs, it is of utmost importance that they are designed to elicit maximum friendliness.

% Previous results have shown that appropriate social behaviors of an IVA can improve the perceived confidence and trust in the IVA.  Friendliness is one of the two dimensions used for the representation of interpersonal relationships, and it refers to the closeness and friendship level between two agents. Thus, in order to perform collaborative tasks with an IVA, modeling their friendliness is important.

 %Therefore, we hypothesize that the consideration of interpersonal social behaviors such as friendliness, will increase the users' sense of social presence with an IVA. 

% Paragraph 2: INTRODUCTION TO MODELING FRIENDLINESS USING FAMILIARITY AND INTERPERSONAL WARMTH Psychology stuff: Friendly--identity.  Familiarity and liking. 
In this paper, we address the problem of automatically generating IVA behaviors that make them appear friendlier by using appropriate cues. Humans voluntarily and involuntarily communicate emotions, moods, and intent via non-verbal cues. Friendliness or warmth is also communicated by non-verbal movement cues such as trajectories, gestures, gaits, etc. Moreover, humans use non-verbal cues such as open gestures, relaxed gaits, and moderate amounts of eye contact to communicate interpersonal warmth~\cite{Reece1962Expressive}. Inspired by these observations in psychology, we investigate methods to generate similar non-verbal movement cues to make our IVAs appear friendlier.
%Since people use these non-verbal cues to convey and perceive friendliness all the time, IVAs that exhibit these nonverbal cues will also appear friendlier and will be more socially-reliable. Additionally, they are also likelier to increase the sense of the presence of users.

% Paragraph 3: INTRODUCTION TO FAMILIARITY IN TERMS OF APPEARANCE, ETC 
% 1.     Familiarity breeds liking? Zajonc     Own race, own family, etc... 
% 2. How do we make something familiar? Faces, body, walking. Uncanny valley--has to look like they are generally in a category. Don't be an alien. 3. Familar digital assistant here's our plan: Taking someone's actual friend. **Could also do someone more like their culture.**
% According to Zajonc REFERENCE, familiarity with an object/person increases its liking for humans. The familiarity can be in terms of appearance and movement. A familiar face, body, walking style can make a virtual agent appear more familiar and as a result, more likable. Appearance of the IVA can thus be used to influence how friendly the users find an IVA.
% \vspace*{0.1in}

\noindent \textbf{Main Results}: We present an approach for modeling Friendly Virtual Agents (FVAs) in an AR environment based on a novel data-driven friendliness model. Our formulation considers three major movement characteristics: gaits, gestures, and gazing. We conduct a perception study and analyze the results to generate a mapping between gaits and perceived friendliness. Moreover, we use psychological characteristics to model the gesture and gazing features that convey friendliness. These movement characteristics are combined and we validate our \textit{Friendliness Model} using an elaborate web-based validation study. 

We use our friendliness model to generate movement characteristics in the form of non-verbal cues for an IVA. We use a Behavioral Finite State Machine (BFSM) to control the gaits, gestures, and gazing cues. Additionally, we augment these movement characteristics with global planning and collision avoidance techniques to perform navigation in the environment. We validate the impact of the generated FVA on users' sense of social presence and their confidence in the FVA's ability to perform designated tasks in an AR environment using a Microsoft HoloLens. Our study results suggest that FVAs provide additional confidence in their abilities to complete standard awareness and influence tasks. Additionally, users experience significantly higher social presence with the FVAs measured using the standard Temple Presence Inventory (TPI) questionnaire.  The novel contributions of our work include:

\begin{enumerate}
    \item Friendliness Model:  We present a novel data-driven mapping between gaits and their perceived friendliness. We also present friendliness models for gestures and gazing derived using psychological characterization. We combine these three friendliness models of gaits, gestures, and gazing to form our overall Friendliness Model (Figure~\ref{fig:friendlinessModel}). We validate the accuracy of this model using a web-based study (Section 4).
    %which evaluates the accuracy of our Friendliness Model. 
    \item We present novel algorithms to generate non-verbal movement characteristics corresponding to the gaits, gestures, and gazing using our friendliness model. Our formulation uses a BFSM, which is updated based upon the state of the environment. We use these algorithms to generate appropriate movement behaviors of IVAs in an AR environment (Section 5).
    %to generate movement characteristics corresponding to gaze, gaits, and gestures using the Friendliness Model. Additionally, a global planning step decides the next intermediate goals for the FVA.
    \item AR Validation Study: We conduct an AR study using a Microsoft HoloLens to evaluate the impact of FVAs on users' sense of social presence and their confidence in the FVA's ability to complete designated tasks in an immersive environment (Figure~\ref{fig:cover}). We hypothesize that FVA will appear friendlier and will increase the user's sense of presence and the confidence in the FVA's abilities. We observe a statistically significant improvement in the perceived friendliness ($5.71\%$ improvement in mean responses) of virtual agents. Our results also suggest that users feel a statistically significant increment in the sense of social and spatial presence with FVAs (Section 6). 
\end{enumerate}

% \begin{itemize}
%     \item An algorithm to generate emotion-specific gaits by modifying emotion features in a neutral gait while maintaining the personality features,
%     \item An approach to create intelligent virtual avatars based on 3D models of known individuals with desired emotional characteristics using non-verbal cues such as gaits, gestures, and gazing, and
%     \item A human evaluation to study the influence of known vs. unknown avatars and emotional characteristics on the users' sense of social presence and confidence in the IVAs.
% \end{itemize}

%The rest of the paper is organized as follows. In Section 2, we review the related work on modeling virtual agents for AR, their movement characteristics and their social behavior. In Section 3, we present relevant background on psychological modeling of friendliness and an overview of our approach to generating FVAs. In Section 4, we present and evaluate the data-driven modeling of friendliness using gaits, gestures, and gazing. We also present our friendliness model and evaluate its perceptual accuracy.  In Section 5, we present details of our algorithm that generates FVAs using the data-driven friendliness models. In Section 6, we present details of the AR validation study.

% In Section 3, we present our approach to model FVAs using nonverbal movement features. We present the details of our human evaluation and discuss the results in Section 4. 

%% file: 2_related.tex
\section{Related Work} % 1 pages
In this section, we give a brief survey of related work on IVAs and their use in AR and mixed reality.   We also  present a brief overview of movement characteristics and social behaviors.

\subsection{Intelligent Virtual Agent Modeling}
Significant research has focused on IVAs and their appearance, behaviors, and movements~\cite{mousas2018effects}. Norouzi et al.~\cite{norouzi2018systematic} surveyed previous work on speech modeling, verbal and non-verbal behavior, physical appearance, and identities of IVAs. Verbal approaches have been used to model different characteristics of speech and dialogues (e.g., emotions~\cite{fraser2018spoken}). Non-verbal approaches corresponding to body posture~\cite{Li2018Effects}, gesture~\cite{Ferstl2018investigating}, facial expression~\cite{lombardi2018deep, nagano2018pagan}, gaze~\cite{Koda2017development}, trajectories~\cite{randhavane2017f2fcrowds,bera2018data}, etc. have been used to model social behavior, affective behavior, and personality traits~\cite{zibrek2018effect}. 
These approaches have also been used for robot navigation among pedestrians~\cite{randhavane2018pedestrian,bera2019emotionally}. 
Other techniques have been proposed to generate different styles of physical appearance (realistic vs. cartoon-like) and evaluate their impact on the sense of presence~\cite{Gerhard2001continuous}. Other approaches reconstructed 3D avatars and poses from videos of real humans for the IVAs~\cite{Lin2016Fast,Tong2012Scanning,li2019technical,facebook}. Beun et al.~\cite{Beun2003Embodied} studied the impact of visual embodiment and anthropomorphism. Normoyle et al.~\cite{normoyle2013effect} studied the effect of posture and dynamics on the perception of emotions of virtual agents. In this paper, we model IVAs with friendliness features using movement characteristics. Our approach uses gaits, gestures, and gazing features and is complimentary to these prior methods.

%Familiarity in terms of appearance (faces and body) increases the liking towards an agent REFERENCE. Previous research has used familiar appearance to influence the perception of an IVA REFERENCE.
\subsection{Movement Characteristics of Virtual Agents}
IVAs have been used to populate AR environments for different applications~\cite{Campbell2005nexus,Amores2014SmartA}. Majority of the research has been focused on designing virtual agents that are aware of the changes in the environment and are capable of influencing the environment~\cite{Chuah2013Exploring}. Prior studies have shown that these abilities have positive effects on co-presence and social presence~\cite{Lee2016the}. IVAs in AR scenarios have been evaluated on standard social presence metrics~\cite{Lombard2009Measuring} and confidence indices~\cite{kim2018does}. In this paper, we design FVAs and evaluate them on similar metrics in an AR environment using a Microsoft HoloLens.
% Vinayagamoorthy2006Building

% ADDRESS: mention that you will restrict to mainly gaze, gestures and gaits.  You can survey other movement characteristics in Section 2 and later mention them in the Future work.

\subsection{Social Behavior of Virtual Agents}
Most previous methods have used Argyle's model of relationship~\cite{argyle1998bodily} for constructing virtual characters~\cite{Bailenson2001Intelligent,Cafaro2012First}. According to this model, interpersonal relationships can be represented using two dimensions: dominance and friendliness. Some researchers have studied dominance relationships of IVAs~\cite{Bee2010Bossy}. In this paper, we focus on modeling the friendliness of IVAs. Friendliness refers to the closeness and friendship levels between two  or more agents. Huang et al.~\cite{Huang2014ModelingAC} modeled friendliness of social robots using non-verbal cues. It is known that friendliness is affected by familiarity and warmth~\cite{fiske2007universal}.  Non-verbal cues such as gaits~\cite{Montepare1998Impressions}, gazing~\cite{argyle1974The}, facial expressions~\cite{Reece1962Expressive}, etc. are used to convey, perceive and communicate warmth. Previous research has used these findings to model warmth of virtual agents. Bergmann et al.~\cite{Bergmann2012Second} used gestures to communicate warmth in virtual agents. Kim et al.~\cite{kim2018does} studied the impact of visual embodiment and social behavior on the perception of IVAs.  Nguyen et al.~\cite{Nguyen2015Modeling} modeled the warmth and competence of virtual agents using a combination of gestures, use of space, and gaze behaviors. Our approach is inspired by these prior methods and we use non-verbal movement characteristics to generate FVAs.

%% file: 3_overview.tex
\section{Overview}
In this section, we introduce our notation and the terminology used in the rest of the paper. We also give a brief overview of the psychological modeling of friendliness. 

\subsection{Notation}
Previous literature differentiates between the virtual representations perceived to be controlled by humans (i.e. avatars) or those perceived to be controlled by computer algorithms (i.e. agents). In this paper, we consider the virtual representation controlled by our algorithm as the FVA. Its visual embodiment is referred to as a 3D model, which consists of a 3D mesh along with a hierarchical skeleton. The hierarchical skeleton is represented by $S$, which contains the positions of all joints relative to their parent joints in a hierarchical tree with the hip joint at its root. The configuration of the skeleton at time $t$ is represented by $conf(S, t)$ and contains the rotation angles of all joints with respect to their parent joints. 

A person's gait or style of walking is a unique feature of their overall movement. A pose $P \in \mathbb{R}^{48}$ of a human is a set of 3D positions of each joint $j_i, i \in \{1,2, ..., 16\}$.  A gait  is a set of 3D poses ${P_1, P_2,..., P_{\tau}}$ where $\tau$ is the number of time steps. Gaits and gestures correspond to time series of configurations of the skeletons. We represent a gait by $g = [conf(S, 0), conf(S, 1), ... , conf(S, n)]$ and a gesture by $m = [conf(S, 0), conf(S, 1), ... , conf(S, n)]$. A set of gaits is represented by $\mathbb{G}$ and a set of gestures is represented by $\mathbb{M}$.
Gaze is an important aspect of human face-to-face interaction and
can be used to increase the behavioral plausibility of the virtual
characters and the overall quality of the immersive virtual experience. We begin by determining if the user  is visible w.r.t the virtual agent and use that information for gaze movement.  We model gazing behavior as a boolean variable that represents eye-contact represented by $\xi$.

% In the initialization step, our algorithm observes the static obstacles in the environment and sets up the coordinate systems with respect to real-world markers.

We use a BFSM to trigger context dependent gestures and gaze for movement cues. The BFSM represents the state of the FVA including its immediate goals and tasks. The BFSM can also include other contexts such as mood and personality. Mathematically, we represent the BFSM by a function $\mathbb{B}: t \times \mathbb{E} \rightarrow \mathbb{I}$, which takes the time and state of the environment $\mathbb{E}$ and outputs a unique BFSM state ID $ i \in \mathbb{I}$. The environment's state contains the positions and dimensions of the static obstacles and the current positions and velocities of the dynamic obstacles. The various movements of an IVA and its trajectory computation is performed using this BFSM.

\subsection{Friendliness}\label{sec:friendlinessInfo}
The notions of friendliness and warmth have been extensively studied in psychology literature. When people interact with others in performing joint tasks, they typically evaluate them on two main dimensions of social cognition: warmth and competence~\cite{fiske2002StereotypeContent,evaluatingcompetence}. The competence dimension concerns whether this agent is able to assist with the task at hand. This question is answered by considering various factors such as ability, intelligence, and reliability. Substantial research in psychology suggests that humans make inferences about other people based upon these factors~\cite{cuddy2011dynamics,cuddy2008dimensions}. Past research on virtual agents or human-computer interaction has also looked at these issues~\cite{Bergmann2012Second,Nguyen2015Modeling}.

The other dimension corresponding to the warmth or friendliness, captures whether people are friendly and well-intentioned. It deals with the most basic interpersonal judgment of whether someone is a friend or foe and whether the other person means to help or harm you. According to Fiske et al.~\cite{fiske2007universal}, warmth is judged before competence, and it has more importance in terms of inducing affective and behavioral reactions. Since psychology literature uses warmth and friendliness in a synonymous manner, we only refer to the term friendliness in the  rest of the paper.
%we are going to use only the term .

To measure friendliness, previous studies~\cite{Bergmann2012Second} used a questionnaire of $18$ items, including \textit{pleasant, friendly, helpful} which have to be assessed on a seven-point Likert scales. Further analysis of this data revealed that the following items tend to measure friendliness with high reliability:  pleasant, sensitive, friendly, helpful, likable, approachable, sociable. In this paper, we also use this scale to measure the friendliness and represent friendliness as a scalar value from $0$ (not at all friendly) to $1$ (very friendly). This scalar value is obtained by averaging the responses to the scale items and normalizing them to $[0, 1]$.

% Friendliness from gaits, gestures, and gazing. Gait representation. Friendliness models. 
% BFSM, state, ID, gestures.

% People make judgments about friendliness using nonverbal cues such as gaits, gestures, gazing, etc. 

% We model the visual appearance of IVAs to convey friendliness using familiarity. To achieve this, we use 3D avatars of the users' friends and well-known celebrities. We also model the movement of IVAs to convey interpersonal warmth between the IVAs and the user. Based on psychological theories, we model nonverbal cues using gait, gazing, and gesture features. In this section, we describe the generation of familiar visual avatars.

% We model the interpersonal warmth between IVAs and the user with nonverbal cues of gaits, gestures, and gazing. These nonverbal cues are formulated using studies in the psychological literature.

%% file: 4_FriendlinessModel.tex
\section{Friendliness Model}
In this section, we describe the friendliness model starting with the friendliness perception study used to obtain a mapping between gaits and their perceived friendliness. Next, we describe the psychological characterizations used to model gestures and gazing features that convey friendliness. We refer to this model of friendliness and non-verbal cues as the \textit{Friendliness Model} in this paper. Finally, we present the details of the web-based validation study used to validate our Friendliness Model.

\begin{figure*}[h]
  \includegraphics[width=0.98\linewidth]{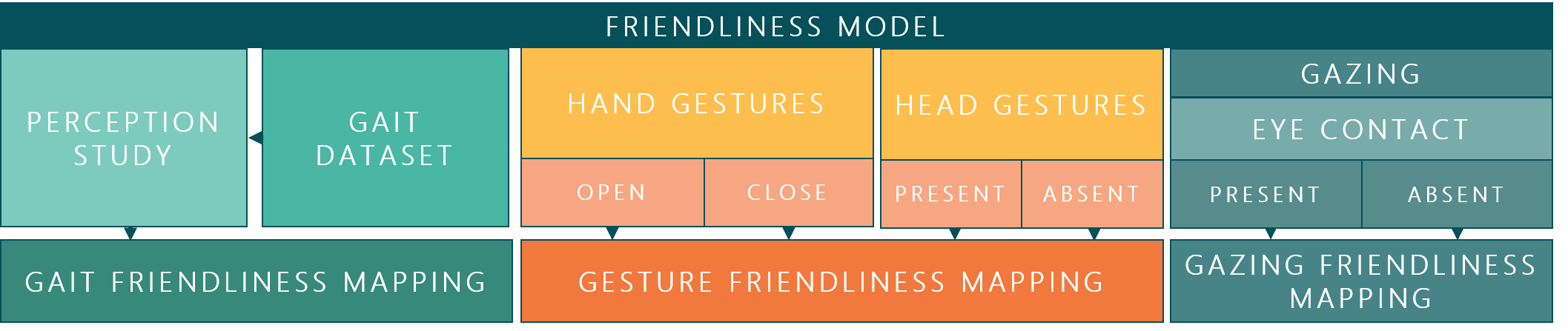}
  \vspace{-10pt}
  \caption{\textbf{Friendliness Model Overview}:  Friendliness can be conveyed using nonverbal cues of gaits, gestures, and gazing. We use data-driven techniques and psychological characterization to formulate a mapping between friendliness and gaits, gestures, and gazing.}
  \vspace{-5pt}
  \label{fig:friendlinessModel}
\end{figure*}

% \begin{figure}[t]
%   \includegraphics[width=0.98\linewidth]{images/warmth.png}
%   \vspace{-10pt}
%   \caption{\textbf{Friendliness and Nonverbal Cues}: Friendliness can be conveyed using non-verbal cues of gaits, gestures, and gazing. We observe that IVAs can appear friendly (left) or unfriendly (right) based on these features. UPDATE: SAY SOMETHING ABOUT THEIR BENEFIT ON THE SOCIAL PRESENCE BASED ON YOUR STUDY RESULTS?}
%   \vspace{-15pt}
%   \label{fig:avatar}
% \end{figure}

\subsection{Friendliness Perception Study for Gaits}\label{sec:perceptionStudy}
This study aimed to obtain a mapping between gaits and their perceived friendliness. 

\subsubsection{Gait Dataset}
\begin{figure}[t]
 \centering
    \includegraphics[width =\linewidth]{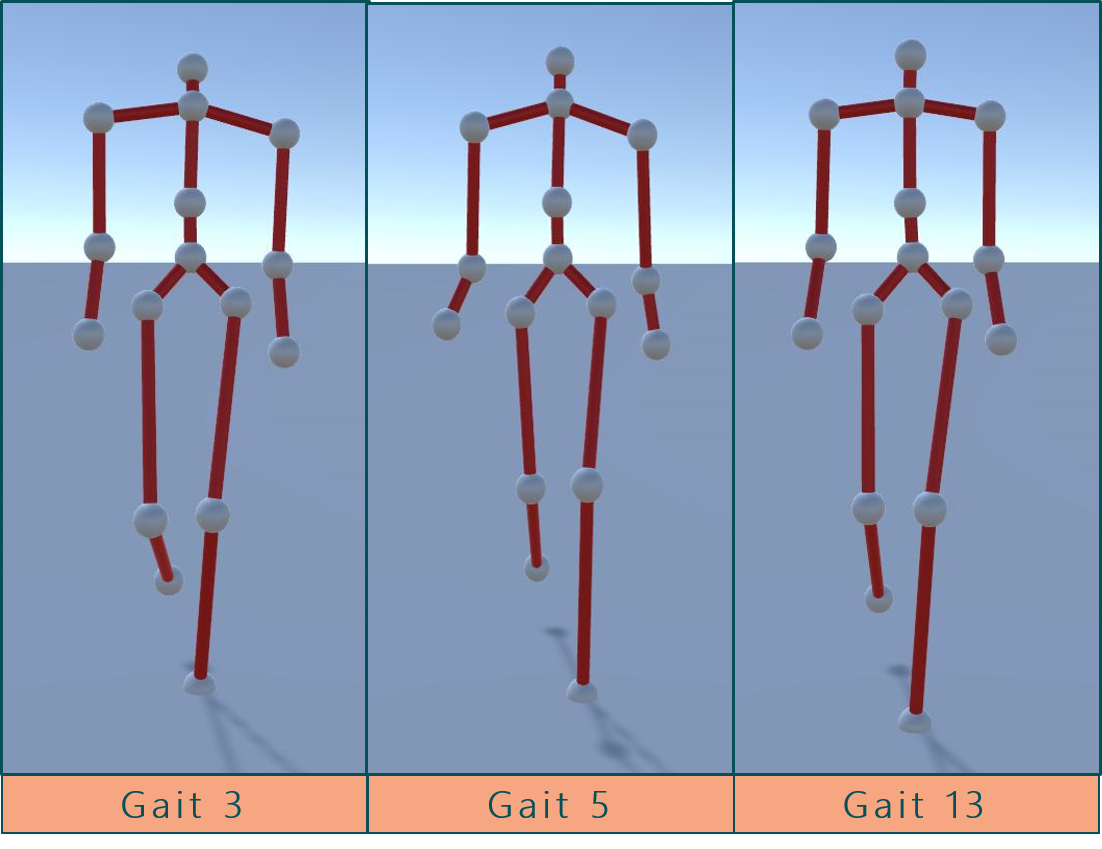}
  \vspace{-20pt}
  \caption{\textbf{Gait Visualizations:} We present sample visualizations of three gaits from a publicly-available, motion-captured gait dataset used for our user-study. We asked the participants to rate each gait video on a $7$-point scale for a friendliness measure (Section 4.1.3). Based on the responses to $49$ gaits from participants, we designed a data-driven model of friendliness and gaits.}
  \label{fig:gaitVideo}
  \vspace{-10pt}
\end{figure}

We designed a data-driven approach and used $49$ gaits from a publicly available dataset of motion-captured gaits~\cite{CMUGait} to compute this mapping. We visualized each gait in the dataset and asked the participants to rate each gait on a friendliness metric. To avoid the appearance of the 3D model affecting the gait's perceived friendliness, we generated the gait visualizations using only a skeleton mesh (Figure~\ref{fig:gaitVideo}). 

\subsubsection{Participants}
We recruited $68$ participants ($35$ female, $33$ male, $\bar{age} = 35.75$) from Amazon Mechanical Turk. We presented a subset of $10$ videos to each participant in a randomized order. The participant could watch each video as many times as he/she wanted and then rate the video on a friendliness measure. For each video, we obtained a minimum of $10$ participant responses.

\subsubsection{Friendliness Measure}
We used a standard friendliness measure (Section~\ref{sec:friendlinessInfo}) with seven components: pleasant, sensitive, friendly, helpful, likable, approachable, sociable. For each component, the participants provided ratings on a $7$-point scale with $7$ indicating a high level of agreement. We also asked whether the participants found the gaits to be unnatural.

\subsubsection{Results}\label{sec:perceptionResults}
We first analyzed the naturalness of gaits and found that the participants found the gaits to be natural (a mean value of $3.87$ for the unnaturalness question). For each gait $g_j \in \mathbb{G}$ in the dataset, participants provided  seven ratings corresponding  to the items of the friendliness measure. We represent a participant response as $r_{j, item} ^ k$ where $j$ is the gait ID, $item$ is one of the items of the friendliness measure, and $k$ is the participant ID. We evaluated the reliability of the measure by computing Cronbach's $\alpha$. We obtained a value of $\alpha=0.922$ indicating that the seven components used for the friendliness measure are reliable. Since social perception is affected by the gender of the observer~\cite{carli1995nonverbal,forlizzi2007interface,kramer2016closing}, we performed a t-test for differences between the responses by male and female participants. We observed that the gender of the participant did not affect the responses significantly ($t = -0.735, p = 0.466$).

We aggregated the participant responses to each gait $g_j$ and obtained the average response to each item $r_{j, item}$ as follows:
\begin{eqnarray}
    r_{j, item} = \frac{\sum_{k=1}^{n_k} r^{k}_{j, item}}{n_k}
\end{eqnarray}
We compute a friendliness value $f_j$ by aggregating the average responses to all items:
\begin{eqnarray}
    f_j = \frac{\sum_{item} r_{j, item}}{7}  \label{eq:gait}
\end{eqnarray}
We normalize the value of $f_j$ to a range $[0, 1]$ where $1$ represents very high friendliness and $0$ represents very low friendliness. 
% We present an histogram of friendliness values obtained for all the gaits in the gait dataset $\mathbb{G}$ in Figure~\ref{fig:friendlinessHistogram}. 

Using the results of this perception study, we obtained a perceived friendliness value $f_j$ for each gait in our dataset $g_j$. We refer to this mapping as the friendliness model of gaits.

% \begin{figure}[t]
%  \centering
%     \includegraphics[width =\linewidth]{images/friendlinessHistogram.png}
%   \vspace{-20pt}
%   \caption{\textbf{Histogram of Friendliness Values of Gaits:} We present a histogram of perceived friendliness values of the gaits in our gait dataset. We can use this friendliness model of gaits to generate gaits with desired friendliness values.}
%   \label{fig:friendlinessHistogram}
%   \vspace{-10pt}
% \end{figure}

\subsection{Friendliness Model for Gestures}
Laban Movement Analysis~\cite{laban1979effort} provides the basis for most research in the area of the description of bodily movement. According to Bergmann et al.~\cite{Bergmann2012Second}, presence of gestures increases the perception of the friendliness of a virtual agent. Using video analysis, Nguyen et al.~\cite{Nguyen2015Modeling} divided hand gestures into two categories: open gestures and closed gestures. These results show that open gestures are perceived as more friendly, and closed gestures are perceived as less friendly. Based on these results, we connect the hand gestures $m_{hand}$ to three levels of friendliness ($f \in [0, 1]$): 
\begin{eqnarray}
    m_{hand} = \begin{cases}
       \text{absent,} &\quad\text{if } f\le 0.33,\\
       \text{closed,} &\quad\text{if } 0.33 \ge f < 0.67, \\
       \text{open,} &\quad\text{if} f \ge 0.67.
     \end{cases}\label{eq:mHand}
\end{eqnarray}
These hand gestures include waving, folding of arms, etc. We also include appropriate head gestures  (e.g., nodding and shaking of the head) for the interaction of our FVA with the user. We connect these head gestures $m_{head}$ to high and low friendliness($f \in [0, 1]$):  
\begin{eqnarray}
    m_{head} = \begin{cases}
       \text{absent,} &\quad\text{if } f < 0.5,\\
       \text{present,} &\quad\text{if } f \ge 0.5.
     \end{cases}\label{eq:mHead}
\end{eqnarray}

\subsection{Friendliness Model for Gazing}
According to previous research, maintaining eye contact is associated with friendliness~\cite{Nguyen2015Modeling}. We control the gazing of the FVA such that the eye contact between the user and the FVA is maintained. This can be computed using visibility computations. In particular, we define the boolean variable $\xi_f$ based on the value of friendliness as follows:
\begin{eqnarray}
    \xi_f = \begin{cases}
       0 &\quad\text{if } f < 0.5,\\
       1 &\quad\text{if } f \ge 0.5,
     \end{cases}\label{eq:fGazing}
\end{eqnarray}
where $0$ represents eye-contact being absent and $1$ represents eye-contact being maintained. 

\subsection{Overall Friendliness Model}
We combine these three friendliness models of gaits (Equation \ref{eq:gait}), gestures (Equation \ref{eq:mHead}), and gazing (Equation \ref{eq:fGazing})  derived using a data-driven approach and psychological characterization using a BFSM. The combined model, after combining and computing the set of gestures $\mathbb{M}$ ($m_{hand}$, $m_{head}$, and $\xi_f$) and gait $\mathbb{G}$  is referred to as the Friendliness Model, which provides appropriate gait, head gesture, hand gesture, and gazing behaviors according to a given friendliness level.

\subsection{Web-based Validation Study}
We conducted a web-based validation study to evaluate our model. This study aimed to validate whether our Friendliness Model was able to model the friendliness of virtual agents (FVA) correctly. 

\subsubsection{Virtual Agents}
\begin{figure}[t]
    \centering
      \includegraphics[width=0.75\linewidth]{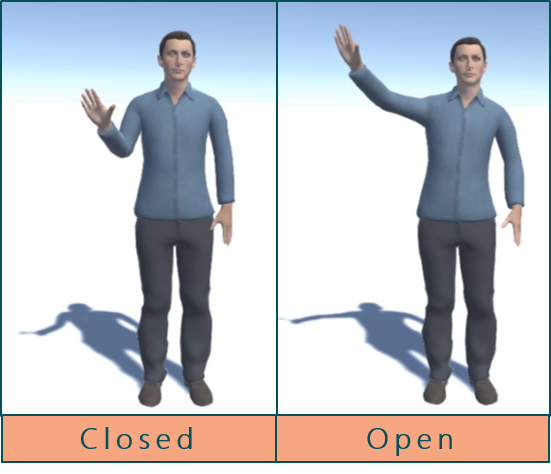}
      \vspace{-10pt}
      \caption{\textbf{Waving Gestures}: We generated videos of virtual agents with nonverbal characteristics corresponding to varying levels of friendliness as predicted by our Friendliness Model. A closed gesture corresponds to a lower friendliness level, whereas an open gesture corresponds to a higher friendliness level. We performed a web-based validation study to evaluate our model using these videos.}
      \vspace{-5pt}
      \label{fig:validation}
\end{figure}

We generated virtual agents with varying non-verbal cues that are modeled using our friendliness model. The virtual agents were visualized with a default male 3D model generated using Adobe Fuse software, which was rigged with a hierarchical skeleton using Adobe Mixamo. Figure~\ref{fig:validation} provides a sample visualization of the virtual agents generated for this study. 

\subsubsection{Nonverbal Movement Characteristics}
For this study, we generated videos of virtual agents with nonverbal characteristics simulated according to the following scenarios:
\begin{itemize}
    \item Walking: This scenario had three videos of the virtual agent walking with three different gaits corresponding to friendliness values $f_{low} = 0.2, f_{medium} = 0.5$, and $f_{high} = 0.9$.
    \item Waving: This scenario had three videos of the virtual agent performing a hand waving gesture with three levels of openness of the gesture.
    \item Nodding: This scenario included a video consisting of two agents standing side-by-side and answering ``yes" to a question. One of the agents performed a nodding head gesture (corresponding to friendliness value $f \ge 0.5$), while the other agent  did not perform any head gesture (corresponding to friendliness value $f < 0.5$).
    \item Shaking: This scenario included a video consisting of two agents standing side-by-side and answering ``no" to a question. One of the agents  performed a head shaking gesture (corresponding to friendliness value $f \ge 0.5$), while the other agent  did not perform any head gesture (corresponding to friendliness value $f < 0.5$).
    \item Gazing: This scenario had a video consisting of two agents standing idly side-by-side.  One of the agents  maintained eye-contact with the user (corresponding to friendliness value $f \ge 0.5$), while the other agent  did not maintain eye-contact (corresponding to friendliness value $f < 0.5$).
\end{itemize}

\subsubsection{Friendliness Measure}
We again used the standard friendliness measure (Section~\ref{sec:friendlinessInfo}).

\subsubsection{Participants}
We recruited $29$ participants ($14$ female, $15$ male, $\bar{age} = 38.82$) on Amazon Mechanical Turk. We presented the videos of the resulting virtual agents to participants in a randomized order. Participants rated each virtual agent on the standard friendliness measure. Therefore, for each virtual agent, we obtained $29$ responses.

\subsubsection{Results}
For each virtual agent, we aggregated the participant responses and obtained friendliness value $f \in [0, 1]$ using the method described in Section~\ref{sec:perceptionResults}. We performed a t-test for differences between the responses by male and female participants. We observed that the gender of the participant did not affect the responses significantly ($t = -1.378, p = 0.184$).

We present the results of this study in Table~\ref{tab:warmthstudy}. The results of this study support that our model of friendliness and nonverbal cues can model the friendliness of virtual agents.

\begin{table*}[t]
\centering
\begin{tabular}{|l|l|l|l|}
\hline
\multicolumn{1}{|c|}{Virtual Agent} & \multicolumn{1}{c|}{Description} & \multicolumn{1}{c|}{f\_pred} & f\_obtained \\ \hline
Gait 1                              & Corresponding to $f_{low}$       & $f_{low} = 0.20$                         & 0.39       \\ \hline
Gait 2                              & Corresponding to $f_{medium}$       & $f_{medium} = 0.50$                         & 0.48        \\ \hline
Gait 3                              & Corresponding to $f_{high}$         & $f_{high} = 0.90$                        & 0.80        \\ \hline
Waving 1                            & Closed waving                    &        $0.33 \le f < 0.67$                      &   0.48          \\ \hline
Waving 2                            & Slightly open waving             & $f \ge 0.67$                 &             0.72 \\ \hline
Waving 3                            & Very open waving                 &  $f \ge 0.67$                             &     0.88        \\ \hline
Nodding 1                        &  No head gesture                                &   $f < 0.5$                           &  0.23           \\ \hline
Nodding 2                       &  Nods head                               &  $f \ge 0.5$                            &     0.89
        \\ \hline
Shaking 1                        &   No head gesture                               & $f < 0.5$                               &   0.30          \\ \hline
Shaking 2                       &   Shakes head                               &  $f \ge 0.5$                                &   0.64         \\ \hline
Gazing 1                         &   Maintains eye-contact                               &     $f \ge 0.5$                             &  0.63           \\ \hline
Gazing 2                        &  Does not maintain eye-contact                                &  $f < 0.5$                              & 0.26            \\ \hline
\end{tabular}
\vspace{5pt}
\caption{\textbf{Results of Web-based Validation Study}: We present the details of the movement characteristics generated and their friendliness values obtained using the web-based validation study. The results of this study indicate that our Friendliness Model correctly models the friendliness using nonverbal movement characteristics corresponding to gaits, gestures, and gazing.}
\label{tab:warmthstudy}
\end{table*}

%% file: 5_FVAs.tex
\section{Friendly Virtual Agents}
In this section, we use our Friendliness Model to generate FVAs. We describe the algorithm to generate movements corresponding to gaits, gestures, and gazing to make an agent appear friendlier in an AR environment.

\begin{figure}[t]
  \includegraphics[width=0.98\linewidth]{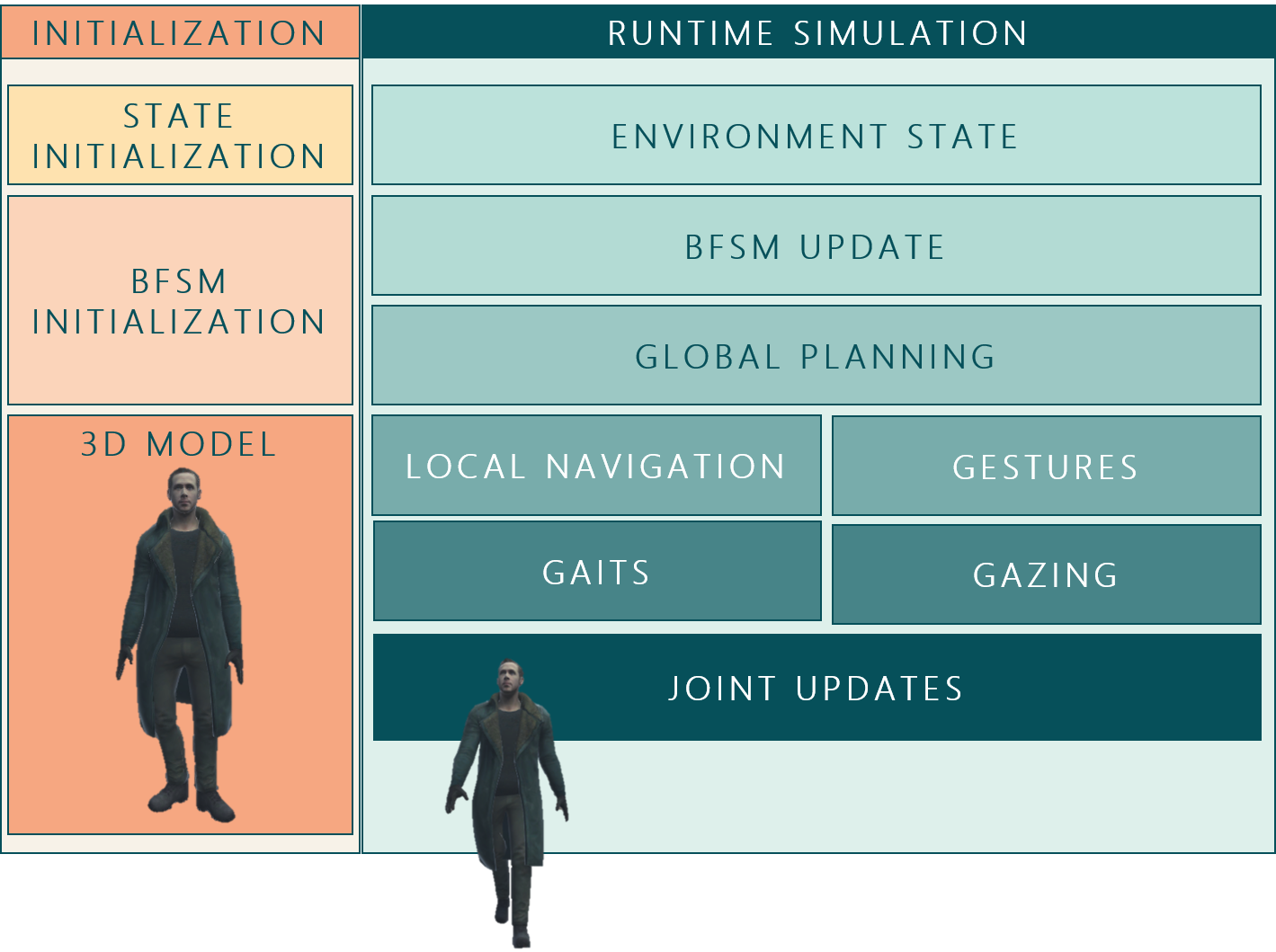}
  \vspace{-20pt}
  \caption{\textbf{FVA Generation}: We highlight our approach to generating a FVA in an AR environment. Our approach uses non-verbal movement characteristics of gaits, gestures, and gazing to convey friendliness. Based on the state-of-the-environment, these features are used to generate the movement of the FVA using a BFSM. We highlight the initialization steps on the left and various components of our runtime system on the right. The BFSM is used to compute each movement cue, which is finally used to update the various joint positions for each FVA. The FVA's movements are visualized in an AR device (e.g., Microsoft HoloLens) using a 3D  human model represented using a mesh and a hierarchical joint-based representation. We use 3D models corresponding to friends and celebrities to evaluate our approach. }
  \vspace{-15pt}
  \label{fig:overview}
\end{figure}

\subsection{Movement Characteristics Generation}
We provide an overview of our end-to-end approach to simulating various movements of an FVA in Figure~\ref{fig:overview}. At the start of the simulation, the environment state and a BFSM are initialized based on the user input and the intended tasks. We use a 3D human model for the representation of the FVA. This model is rigged using an automatic rigging software and a hierarchical skeleton is associated with appropriate joint values. 
% We use the joints of this hierarchical skeleton to update the visualization of the agent. 
% of a user's friend or a celebrity with a positive impression. We choose these familiar 3D model for the FVA because according to psychology research familiarity increases friendliness.

At runtime, we update the BFSM state according to the state of the environment and whether the FVA has completed the assigned tasks. Based on the updated state, a global planning step is used to compute the next intermediate goal position and configuration for the FVA. Next, we compute velocities for local navigation (if it needs to change location in the real world) and decide gaits, gestures, and gazing behaviors according to the BFSM state. We then update the joint positions and angles of the FVA's hierarchical skeleton using the computed gaits, gestures, and gazing features. We visualize the FVA using the selected 3D model in an AR device such as the Microsoft HoloLens. We describe these components in detail below.

\subsection{Behavioral Finite State Machine}
We use a BFSM to represent the behavioral state of the FVA and control its behavior. We assume that the environment consists of static obstacles and dynamic obstacles. We also assume that the locations of these static obstacles and their dimensions can be reliably estimated. Additionally, dynamic obstacles can also be tracked, and their positions and velocities can be reliably determined at each time step. These dynamic obstacles may include humans in the real world (including the user) and virtual agents. During the initialization step, the environment's state is initialized with positions and dimensions of the static obstacles and the current positions and velocities of the dynamic obstacles. We represent this state of the environment by $\mathbb{E}$. Using the environment state and the intended task, a BFSM is initialized and the starting state of the BFSM is computed for a given task.  

At runtime, we update the state of the BFSM based on the environment state and determine the FVA's behavior within the context of the current task. The BFSM also computes a goal position $\vec{g}_i$ for each virtual agent. If there are obstacles or other virtual agents in the environment, we make sure that the resulting movements are collision-free.

\subsection{Global and Local Navigation}
If the goal position of virtual agents is different from their current position, then a navigation algorithm is used to compute the trajectory to the new position. We utilize the multi-agent simulation framework, \textit{Menge}~\cite{curtis2016menge}, to implement our local navigation algorithm. In this framework, a global navigation step breaks down the goal positions into intermediate goals that avoid collisions with the static obstacles in the environment. Moreover, a local navigation step provides collision-free navigation to the intermediate goals while avoiding collisions with dynamic obstacles, which include other virtual agents and users' avatars, using a reciprocal collision avoidance approach. This computes a trajectory in the 2D plane. We ensure that no collisions are introduced due to other joint movements.
% FVAs perform collision avoidance and show the awareness of the environment.

\subsection{Gait Generation}
When the agent is walking towards its goal position, as determined by the BFSM and the local navigation algorithm, its walking style is modeled using its gait $g_{des}$. Based on the input friendliness value $f_{des}$, we use the Friendliness Model to obtain the gait $g_{des}$ from the gait dataset $\mathbb{G}$ as follows:
\begin{eqnarray}
    g_{des} = g_j \mid j = \min_{j = 1}^{n}\ abs(f_j - f_{des})\ \land\ g_j \in \mathbb{G},
\end{eqnarray}
where $n$ is the number of gaits in $\mathbb{G}$ and $abs(\cdot)$ represents the absolute value. This way we ensure that the gait of the FVA has the desired friendliness based on our model as presented in Section 4.1.

\subsection{Gestures}
We associate the BFSM state with gestures based on a mapping $\mathbb{GE} : \mathbb{I} \rightarrow \mathbb{M}$, where $\mathbb{M}$ represents a set of gestures $m_{j}\ (j = 1...n)$ and $\mathbb{I}$ represents the set of BFSM states. This mapping is based on the BFSM context and on the use of open gestures to convey a high level of friendliness. For example, the FVA performs a waving gesture when the FVA exits the scenario or nods its head while indicating agreement with the user. Given the appropriate hand and/or head gestures $m_{hand}$ and $m_{head}$, we use the input friendliness value $f_{des}$ and the Friendliness Model formulation (Equations~\ref{eq:mHand} and \ref{eq:mHead}) to decide whether to use the open or closed gesture.

\subsection{Gazing}
In addition to friendliness, maintaining eye-contact or gazing also depends on the environment state (e.g., eye-contact should not be maintained while walking away~\cite{narang2016pedvr}). Therefore, we associate a BFSM state to decide whether to maintain eye-contact or not $\mathbb{GA} : \mathbb{I} \rightarrow \xi_{BFSM}$ where $\xi_{BFSM}$ is a boolean variable. We combine this function with the gazing behavior decided by the Friendliness Model $\xi_f$ (Equation~\ref{eq:fGazing}) and compute a variable $\xi$:
\begin{eqnarray}
    \xi =  \xi_f\ \land\ \xi_{BFSM},
\end{eqnarray}
where $\xi = 0$ represents that the eye-contact is absent and $\xi = 1$ conveys that the eye-contact is being maintained. 

\begin{figure}[t]
  \includegraphics[width=0.98\linewidth]{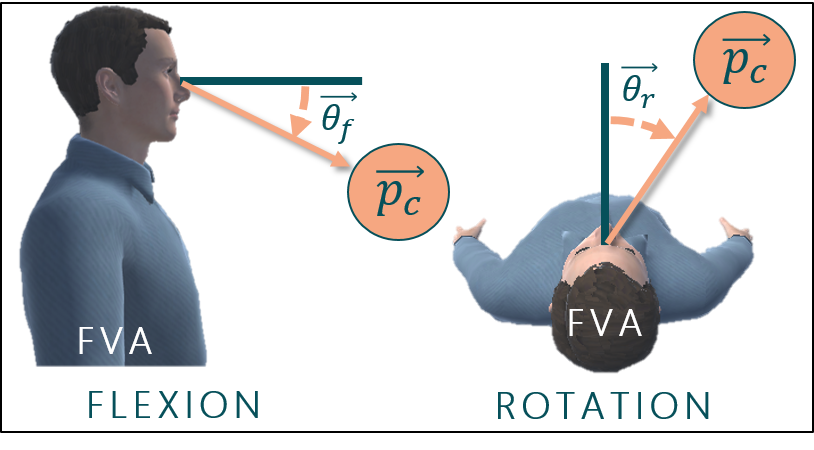}
  \vspace{-10pt}
  \caption{\textbf{Gazing Cues}: Friendliness can be conveyed using nonverbal cues associated with gazing. We model gazing features by computing the neck flexion and rotation angles such that the FVA maintains eye-contact with the user (represented by the rendering camera, $\vec{p}_c$).}
  \vspace{-15pt}
  \label{fig:avatar}
\end{figure}

We use the FVA's 3D model to maintain eye-contact during a movement or task. Specifically, we manipulate the flexion and rotation angles of the FVA's neck joint (Figure~\ref{fig:avatar}). We use the 3D positions of the user's eye (represented by the rendering camera, $\vec{p}_c$)) and the position of the FVA ($\vec{p}_{FVA}$) to compute the flexion $\theta_f$ and rotation $\theta_r$ angles: 
\begin{eqnarray}
    \theta_{f} &=& asin\Big(\frac{\vec{p}_{c, z} - \vec{p}_{FVA, z}}{||\vec{p}_{c} - \vec{p}_{FVA}||}\Big),  \\
    \theta_{r} &=& asin\Big(\frac{\vec{p}_{c, x} - \vec{p}_{FVA, x}}{||\vec{p}_{c} - \vec{p}_{FVA}||}\Big).
\end{eqnarray}
We can also check the line-of-sight between the agent and the user using visibility queries.
For this computation, we assume that the FVA is facing the x-axis and the z-axis points vertically up. Other orientations of the FVA can be similarly modeled by adding a coordinate transform in the computation. 

% In the BFSM's states where the FVA is walking, we use gait features to compute the orientations of all the body joints of the FVA. The agent maintains eye-contact using our computation of neck flexion and rotation angles when the agent is idle or not looking at another target specified by the BFSM state.

%% file: 6_userStudy.tex
\section{Results and AR Validation Study}
In this section, we describe the implementation of FVA in an AR environment using a Microsoft HoloLens. We also present the details of our validation study conducted in an AR scenario to evaluate the benefits of FVA in terms of social presence.

Our algorithm is implemented on a Windows 10 Desktop PC with an Intel Core i7-7820HK CPU at 2.90 GHz, Nvidia GeForce TitanX graphics card, and 16 GB of RAM. All the behavior modules run on the CPU, and we use the NVIDIA GPU for rendering. In practice, our system can render virtual agents at interactive rates of \texttildelow70 FPS. Furthermore, we can handle a few tens of agents at interactive rates (i.e. at frame rates more than $30$ FPS). As a user walks around the AR environment or performs the tasks, we control the friendliness of IVAs using our approach based on their gaits, gestures, and gazing. To evaluate the perceptual benefits, we also conducted an AR Validation study.
% We present the details of the experiment we conducted to investigate the effects of familiarity and interpersonal warmth on the perception of IVAs during social interaction. ~\cite{kim2018does}

\subsection{Experiment Goals}
The goal of this experiment was to evaluate the impact of our FVAs in terms of users' sense of social presence with and their confidence in the FVA within an AR environment.

\subsection{Participants}
Based on prior studies~\cite{kim2018does} and our initial pilot tests, we decided to recruit $20$ participants, which proved to be sufficient to show significant effects in our experiment. We recruited $10$ female and $10$ male participants for our experiment ($\bar{age} = 27.75$) from a university community. All the participants had a correct or corrected-to-normal vision. Three participants wore glasses during the experiment. None of the participants reported any history of discomfort with the HoloLens. 

\subsection{Virtual Agents}
For this experiment, we used two types of 3D models to represent the virtual agents. One of the models, \textit{Ryan}, was a commercially-available 3D model of a celebrity. The other model, \textit{John}, was designed and rigged using Adobe Fuse and Adobe Mixamo. We used two different models to account for variation in appearance. Previous research suggests that females appear friendlier than males~\cite{abele2003dynamics}. We control this gender effect by conducting the study with only male avatars. We generated virtual agents using two methods:

\begin{itemize}
    \item Default: A virtual agent with a default gait (friendliness value of $f = 0.52$) and with gestures and gazing features absent. We refer to this agent as the \textit{Default} agent in the rest of this section.
    \item FVA: A virtual agent generated using our Friendliness Model. We used a gait with friendliness value of $f=0.97$. The agent performed appropriate gestures, including head nodding and waving. The agent also maintained eye-contact with the participant (represented by the rendering camera).
\end{itemize}

Based on these two 3D models and two methods, we obtained four variations of virtual agents in our AR environment: \textit{DefaultRyan, FVARyan, DefaultJohn, FVAJohn}.

\begin{table*}[t]
\resizebox{\textwidth}{!}{%
\begin{tabular}{|p{10pt}|p{30pt}|p{135pt}|p{150pt}|p{120pt}|}
\hline
\multicolumn{1}{|c|}{Task} & \multicolumn{1}{c|}{Description} & \multicolumn{1}{c|}{Participant Command}                                  & \multicolumn{1}{c|}{Agent Acceptance Response}                                   & \multicolumn{1}{c|}{Agent Completion Response}                   \\ \hline
A1                            & See                              & Please check if anyone is in the adjacent room.                           & Okay! I am checking if anyone is in the adjacent room right now.                 & There are a few people in the adjacent room.                     \\ \hline
A2                            & Hear                             & Please check if it is quiet enough to perform the experiment.             & Okay! I am checking if it is quiet enough to perform the experiment.             & It is quiet enough to perform the experiment.                    \\ \hline
A3                            & Feel                             & Please check if the temperature is high enough to conduct the experiment. & Okay! I am checking if the temperature is high enough to conduct the experiment. & The temperature is high enough to conduct the experiment.        \\ \hline
I1                            & Physical                         & Please close the adjacent room's other entrance.                          & Okay! I am closing the adjacent room's other entrance.                           & I closed the adjacent room's other entrance.                     \\ \hline
I2                            & Social                           & Please tell someone that the experiment will end in 15 minutes.           & Okay! I am telling someone that the experiment will end in 15 minutes.           & I told someone that the experiment will end in 15 minutes.       \\ \hline
I3                            & Social Critical                  & Please tell someone that I am not feeling well.                           & Okay! I am telling someone that you are not feeling well.                        & I told someone that you are not feeling well.                    \\ \hline
I4                            & Digital                          & Please turn off the audio and video recording in the adjacent room.       & Okay! I am turning off the audio and video recording in the adjacent room.       & I turned off the audio and video recording in the adjacent room. \\ \hline
\end{tabular}%
}
\vspace{5pt}
\caption{\textbf{Tasks Performed in Our AR Validation Study}: Participants in an AR environment used a Microsoft HoloLens. Each participant was asked to give commands to the virtual agent to perform three awareness tasks and four influence tasks. The awareness tasks (A1, A2, and A3) were related to the agent's ability to sense the real world whereas the influence tasks were (I1, I2, I3, and I4) were related to the agent's ability to influence it. In each task, after the participant gave the command, the agent responded with an acceptance response and proceeded to perform the task. After completing the task, the agent provided the completion response. After each task, the participant answered a question that measured the participant's confidence in the virtual agent's ability to complete the task. After the completion of the last task, the agent responded ``Bye Bye" and the simulation ended. The participants then answered a friendliness questionnaire (Section~\ref{sec:friendlinessInfo}) and a subset of the Temple Presence Inventory to measure the social presence. } 
\vspace{-10pt}
\label{tab:tasks}
\end{table*}

\subsection{Study Design}
To compare the \textit{FVA} with the \textit{Default} agent, we wanted to let the participants interact with two types of agents directly. Therefore, we used a within-subjects design where each participant interacted with all four variations of the virtual agents. We presented the virtual agents in a randomized order to account for the order effects. 

\subsection{Procedure}
Our experiment and evaluation were conducted in a laboratory setting. After entering the laboratory, the participants were informed about the study and the procedure. After consenting to participate in the study, the participants sat down in a chair and wore the Microsoft HoloLens HMD.  We used a physical room-like experiment with an area of $4.0 m \times 4.0 m$. There was a door in front of the participant's chair that led to another room, referred to as the ``adjacent room" in the rest of the paper.

Each participant performed four sessions corresponding to the four virtual agents in a randomized order. In each session, the participants and the agent interacted as part of a scripted pseudo-real story to perform some tasks based on the participant's verbal commands. This story was designed based on previously used methods to evaluate the confidence and social presence of virtual agents in an AR environment~\cite{kim2018does}. The participants were provided with instructions about how to progress through the story using an iPad. The iPad also served as an interface for answering the questionnaire for the users.

\subsection{Interaction Scenario}
The main story started with an \textit{introduction} stage, where the virtual agent introduced itself as the ``lab assistant" who will help the participant conduct an experiment. In this stage, the virtual agents \textit{FVARyan} and \textit{FVAJohn} maintained eye-contact with the participants, whereas \textit{DefaultRyan} and \textit{DefaultJohn} did not maintain eye contact. This helped us evaluate the benefits of gazing. 

After the introduction stage, we begin the main story involving the participant asking the virtual agent (lab assistant) to perform some tasks in the adjacent room. Participants performed seven tasks per session related to the virtual agent's awareness of the environment (three tasks) and its influence on the environment (four tasks). We list the tasks in Table~\ref{tab:tasks}. The awareness tasks (A1, A2, and A3) are related to the agent's ability to sense the real world. Specifically, they involve the agent's ability to see, hear, and feel the real physical world. These abilities are analogous to humans' natural abilities to sense the world around them. For the virtual agents, these abilities can be realized with cameras, microphones, and thermometers, respectively. 

The influence tasks (I1, I2, I3, and I4) are related to the virtual agent's ability to influence the real world. Specifically, they involve the agent's abilities in terms of physical, social, social critical, and digital influence. These abilities can be realized by devices such as smart-doors~\cite{morris2013smart} and speakers. In our experiment, we did not implement the actual functionality and used pre-defined agent responses. 

Within each of the seven tasks, the participant first gave a verbal command to the agent according to the instructions provided on the iPad. To avoid errors arising from speech processing, we used a human-in-the-loop mechanism where a human operator triggers the agent acceptance response after hearing the participant's commands. The agent responded to the participant via text. After providing the acceptance responses, the agent navigated to the adjacent room and performed the task. To create an impression of performing the task, the virtual agent stood facing away from the participant for five seconds. The agent then walked back and stood in front of the participant and provided a completion response. In the case of \textit{FVARyan} and \textit{FVAJohn}, the FVA also performed a nodding gesture while giving the completion response. After receiving the completion response, participants answered a specific task-related item using the iPad, and the agent awaited the next task.  After the completion of the last task, the agent responded ``Bye Bye" and the entire session of the user in the AR environment ended. In the case of \textit{FVARyan} and \textit{FVAJohn}, the FVA also performed a hand waving gesture accompanying the ``Bye Bye" response.

\subsection{Measures}
We administered four types of measures: friendliness, awareness, influence, and social presence.

\subsubsection{Friendliness}
We again used the standard friendliness measure (Section~\ref{sec:friendlinessInfo}).

\subsubsection{Awareness and Influence}
After completing each of the awareness tasks (A1, A2, and A3) and the influence tasks (I1, I2, I3, and I4), the participants rated their confidence in the abilities of the agent to complete the task. We asked questions -- ``How confident are you that the agent was able to ...?" with modifications according to the specific task. These questions were meant to assess the impact of our algorithms on the participants' confidence in the abilities of the agent based on the participants' immediate reactions to the virtual agents.

\subsubsection{Social Presence}
For each of the four virtual agents, after completing the seven tasks, the participants answered a subset of the Temple Presence Inventory (TPI)~\cite{Lombard2009Measuring}. TPI is a questionnaire designed to assess co-presence and social presence with virtual agents. For our task, we use a slightly modified subset of the original TPI questionnaire to measure the social presence in our AR scenario. Specifically, we considered the following subscales:
\begin{itemize}
    \item Social Presence: How much does the participant feel that the virtual agent is in the same space as them and how well does the participant and the virtual agent interact/communicate?
    \item Spatial presence: How much does the participant feel that the virtual agent has come to the place where they are co-located and how much does the participant feels that they can reach out and touch the virtual agent?
    \item Social richness: How much does the participant perceive the virtual agent as sociable, warm, sensitive, personal, or intimate?
    \item Engagement: How immersive or exciting is the interaction with the virtual agent so that the participant feels deeply involved in the interaction?
\end{itemize}

\subsection{Hypotheses}
We propose the following hypotheses:
\begin{itemize}
    \item \textbf{H1}: FVA will appear friendlier to the participants compared to the Default agent.
    \item \textbf{H2}: Participants will exhibit more confidence in the FVA's awareness of the real world compared to the Default agent. 
    \item \textbf{H3}: Participants will exhibit more confidence in the FVA's ability to influence the real world compared to the Default agent.
    \item \textbf{H4}: Participants will feel a stronger sense of social presence with the FVA compared to the Default agent.
\end{itemize}

\begin{table}[t]
\centering
\begin{tabular}{|l|c|c|c|c|}
\hline
\multirow{2}{*}{} & \multicolumn{2}{c|}{Default} & \multicolumn{2}{c|}{FVA} \\ \cline{2-5} 
                  & Mean          & SD           & Mean        & SD         \\ \hline
Friendliness      & 4.707         & 1.216        & 5.050       & 0.915      \\ \hline
Awareness         & 5.075         & 1.239        & 5.308       & 1.082      \\ \hline
Influence         & 4.938         & 1.149        & 5.056       & 1.136      \\ \hline
Social Presence   & 4.006         & 1.073        & 4.247       & 1.015      \\ \hline
Spatial Presence  & 3.115         & 0.723        & 3.440       & 1.045      \\ \hline
Social Richness   & 4.346         & 1.232        & 4.729       & 0.969      \\ \hline
Engagement        & 4.266         & 1.443        & 4.446       & 1.476      \\ \hline
\end{tabular}
\vspace{5pt}
\caption{\textbf{Mean Responses to Measures}: We present the participants' average responses to each of the measures for \textit{Default} and \textit{FVA} agents. \textit{FVA} agents have higher means across all the measures compared to the \textit{Default} agents.}
\label{tab:ARmeans}
\end{table}

\begin{table}[t]
\centering
\begin{tabular}{|l|c|c|}
\hline
                 &   $\chi^2$    &  $p$     \\ \hline
Friendliness     & 4.000 & 0.046 \\ \hline
Awareness        & 0.030 & 0.862 \\ \hline
Influence        & 0.118 & 0.732 \\ \hline
Social Presence  & 4.000 & 0.046 \\ \hline
Spatial Presence & 6.125 & 0.013 \\ \hline
Social Richness  & 2.189 & 0.139 \\ \hline
Engagement       & 1.485 & 0.223 \\ \hline
\end{tabular}
\vspace{5pt}
\caption{\textbf{Friedman Test}: We present the $\chi^2$ and $p$-values for the Friedman test performed for the  \textit{Default vs. FVA} comparison for different measures and subscales.}
\label{tab:ARFriedman}
\end{table}

% \begin{table}[t]
% \centering
% \begin{tabular}{|l|l|l|}
% \hline
% \multicolumn{1}{|c|}{} & \multicolumn{1}{c|}{$\chi^2$} & \multicolumn{1}{c|}{$p$} \\ \hline
% Friendliness           & 3.200                 & 0.074                 \\ \hline
% Awareness              & 0.330                 & 0.564                 \\ \hline
% Influence              & 0.620                 & 0.433                 \\ \hline
% Social Presence        & 1.190                 & 0.275                 \\ \hline
% Spatial Presence       & 2.880                 & 0.896                 \\ \hline
% Social Richness        & 1.580                 & 0.209                 \\ \hline
% Engagement             & 1.810                 & 0.178                 \\ \hline
% \end{tabular}
% \vspace{5pt}
% \caption{\textbf{Friedman Test}: We present the $\chi^2$ and $p$-values for the Friedman test performed for the  \textit{Default vs FVA} comparison for different measures and subscales.}
% \label{tab:ARFriedman}
% \end{table}

\subsection{Results}
We chose a familiar celebrity character (Ryan) and a default character (John) because familiarity affects friendliness~\cite{fiske2007universal}. Our findings corroborate these psychological findings (mean friendliness for $John = 4.921, Ryan = 5.179, p = 0.026$). For further analysis, we combined participant responses to Ryan and John to show the contribution of our friendliness algorithm, which is agnostic to the familiarity with the visual appearance.

We performed a t-test for differences between the responses by male and female participants. We observed that the gender of the participant did not affect the responses significantly ($t = -0.196, p = 0.845$).

We present average participant responses for the four measures (including  subscales) for the \textit{Default} and the \textit{FVA} agents in Table~\ref{tab:ARmeans}. Since the participant questionnaire responses are of an ordinal data type, we used non-parametric Friedman tests to compare between responses for the two types of virtual agents. For this test, the method used to generate the virtual agent (\textit{Default vs. FVA}) is the independent variable, and the participant response is the dependent variable. We present the test statistic $\chi^2$ and the p-value $p$ for this test in Table~\ref{tab:ARFriedman}. The results of this test reveal significant differences between the two methods for friendliness, social, and spatial presence.

\subsection{Discussion}
We now discuss these results in more details.

\subsubsection{Friendliness}
Our hypothesis H1 was concerned with the friendliness of FVA in an AR environment compared to the Default agent. We observed that there was a statistically significant difference between the two methods for the friendliness comparison $(\chi^2 = 4.000, p = 0.046)$ with FVA reporting a $5.71\%$ higher mean compared to the Default agent. These results support the hypothesis H1.

\subsubsection{Awareness and Influence}
Our hypothesis H2 was concerned with the confidence in the abilities of the FVA to sense the real world compared to the Default agent. We observed that participants reported a $3.89\%$ higher mean compared to the default agent, however this difference was not significant $(\chi^2 = 0.030, p = 0.862)$.

% We observe that there is a statistically significant difference between the two methods for the awareness task with FVA reporting $8.17\%$ higher mean compared to the Default agent. These results support the hypothesis H2.

Our hypothesis H3 was concerned with the confidence in the abilities of FVA to influence the real world compared to the Default agent. We observed that participants reported a $1.98\%$ higher mean compared to the default agent, however this difference was not significant $(\chi^2 = 0.118, p = 0.732)$.

These results indicate that our Friendliness Model increases the users' confidence in the virtual agent's abilities to sense and influence the real world. However, the results did not significantly support hypotheses H2 and H3, possibly because these questions also attempted to evaluate the competence, which is an independent dimension of social cognition and is not affected by friendliness~\cite{fiske2007universal}.

\subsubsection{Social Presence}
The statistically significant differences  $(\chi^2 = 4.000, p = 0.046)$ between the two types of agents, FVA and Default, for social presence support our hypothesis H4 that the participants indeed feel a stronger sense of social presence with the FVA compared to the Default agent. 

We also observed statistically significant improvement in the spatial presence $(\chi^2 = 6.125, p = 0.013)$. However, the improvement was not significant for social richness and engagement subscales. In optional feedback, the participants reported that they sometimes forgot that the FVA was a virtual agent and they wanted to wave back when the FVA performed the waving gesture.

%% file: 7_Conclusion.tex
\section{Conclusions, Limitations, and Future Work}
In this paper, we presented a novel Friendliness Model based on three non-verbal movement characteristics corresponding to gaits, gestures, and gazing. Our data-driven model was generated with a user study, and we validated its benefits using a web-based study. Based on our model, we present algorithms to interactively generate the movement characteristics of a virtual agent (FVA) to make it appear friendly. We performed an extensive AR validation study using Microsoft HoloLens to evaluate the benefits of our model. Our study results indicate that FVAs cause a statistically significant increase in the sense of social presence in an AR environment. Our study results indicate that FVAs cause a statistically significant increase in the sense of social and spatial presence in an AR environment ($5.42\%$ improvement in the mean participant response). Our work has both methodological and theoretical implications for psychology research, especially in terms of evaluating social psychological methods. Our study uses multiple dynamic, naturalistic channels of stimuli to reveal the integration of multiple social cues on social judgments (i.e. friendliness) and explores the basic psychological processes used by the people to perceive friendliness in others.

Our approach has some limitations. In our experiment, we did not implement the actual functionality for sensing and influencing the environment and used pre-defined agent responses. Additionally, we only consider a subset of non-verbal and movement cues to control the behavior of a virtual agent. We used a pre-determined set of gestures and a mapping between BFSM states and the gestures. Many other components also govern the perception of friendliness and warmth of virtual agents, including speech, facial expressions, and other characteristics of body language as well as cultural difference. Age and gender have been shown to affect the perception of friendliness~\cite{abele2003dynamics,carli1995nonverbal,harper1985power}, however in the AR validation study we did not consider these variations.

There are many avenues for future work in addition to overcoming these limitations. Learning approaches have been developed for character motion synthesis~\cite{holden2016deep,holden2017phase} and speech-based automatic gesture generation~\cite{Ferstl2018investigating,hasegawa2018evaluation}. In the future, we want to combine our approach with these deep learning- and controller-based methods. We want to extend our approach to generate gestures with desired levels of friendliness automatically based on these learning-based approaches. Non-verbal movement cues also affect other dimensions of interpersonal trust and we would like to evaluate them. We also want to provide virtual agents with sensing and influencing capabilities using cameras, microphones, thermometers, smart-doors, speakers, etc. We want to develop an approach to designing virtual agents for AR that appear friendly as well as competent.